\documentclass[12pt]{article}

\usepackage{amsmath}
\usepackage{amsthm}
\usepackage{amssymb}
\usepackage{color}

\def\innerprod(#1,#2){{\left<#1\,,\,#2\right>}}

\def\qquadand{\qquad\text{and}\qquad}

\def\iv{i_V}
\def\iu{i_U}

\def\dual#1{{\widetilde{#1}}}
\def\dualv{{\dual{V}}}
\def\dualu{{\dual{U}}}

\def\ivF{\iv F}
\def\ivstarF{\iv{{\star}}F}
\def\iuF{\iu F}
\def\iustarF{\iu{{\star}}F}
\def\ivG{\iv G}
\def\ivstarG{\iv{{\star}}G}

\def\Me{{\mathbf e}}
\def\Mb{{\mathbf b}}
\def\Md{{\mathbf d}}
\def\Mh{{\mathbf h}}

\def\PermDE{{\zeta^{\textup{de}}}}
\def\PermDB{{\zeta^{\textup{db}}}}
\def\PermHE{{\zeta^{\textup{he}}}}
\def\PermHB{{\zeta^{\textup{hb}}}}

\def\PermTen{{Z}}

\def\GenPoyn{{\mathbf s}}
\def\SEM{ stress-energy-momentum }

\def\EM{ electromagnetic }

\def\ee{\,\epsilon_0\,}
\def\cc{\, c \,}

\begin{document}
\begin{titlepage}
\title{ The Covariant Description of
Electromagnetically
   Polarizable Media}
\author{ T. Dereli\footnote{E.mail: tdereli@ku.edu.tr}
\\{\small Department of Physics, Ko\c{c} University}
\\{\small 34450 Sar{\i}yer,  \.{I}stanbul, Turkey}\\
\\J.  Gratus\footnote{E.mail: j.gratus@lancaster.ac.uk} \\
{\small Department of Physics, Lancaster University and the Cockcroft Institute}\\
{\small Lancaster LA1 4YB, UK},\\
\\R. W. Tucker\footnote{E.mail: r.tucker@lancaster.ac.uk} \\
{\small Department of Physics, Lancaster University and the
Cockcroft Institute}\\ {\small Lancaster LA1 4YB, UK}}

\maketitle


\vspace{0.4cm} {\small

 \centerline{{\bf Abstract}}

\noindent The form of the phenomenological stress-energy-momentum
tensor for the electromagnetic field in a class of inhomogeneous,
anisotropic magneto-electric media is calculated from first
principles, leading to a coherent understanding of the
phenomenological stresses and energy-momentum exchanges induced by
electromagnetic interactions with such matter in terms of a fully
relativistic covariant variational framework.

}

\end{titlepage}
\section{Introduction}

The natural mathematical tool for describing  in a covariant
manner the response of matter to excitation by external fields is
 the total \SEM tensor
involving  matter and  fields. Early suggestions by Minkowski
\cite{mink} and Abraham \cite{abr} for the structure of its \EM
component in simple media initiated a long debate involving both
theoretical and experimental contributions that continues to the
current time (see e.g.\,\cite{peierls1}, \cite{peierls},
\cite{gordon}, \cite{brevik}, \cite{loudon}, \cite{ob_hehl},
\cite{ob_hehl_post}, \cite{feigel}, \cite{bowyer},
\cite{cambell}). Disputes developed because different proposals to
describe the interaction of light with media  were adopted and
subsequent experiments were unable to resolve their conflicting
predictions. Although it is widely recognised today that this
controversy is an argument about definitions \cite{feld},
\cite{mikura}, \cite{nelson}, \cite{israelstewart} and that the
relative merits of alternative definitions are undecidable without
a complete (experimentally verifiable) covariant description of
relativistic continuum mechanics for matter and fields, the
absence of a compelling derivation of a total relativistic
stress-energy-momentum tensor that can effectively model the
phenomenological response of moving media to electromagnetic
fields  is surprising.

If  \SEM tensors are defined as appropriate variational
derivatives of an action functional,  a mathematically precise and
physically cogent derivation of a symmetric
 tensor can be effected.
In this letter we report that  both the original phenomenological
proposals by Minkowski and Abraham can be viewed in this context
and correspond to different choices for the response of a linear
constitutive tensor to gravitation. Since it is known that the
polarisation and magnetisation of most continua depend on their
state of motion the choice made by Abraham (originally in the
context of inertial motions and zero gravitation) probably offers
a more effective contribution to the total stress-energy-momentum
tensor. From this perspective it is also argued that if spatial
and temporal dispersion are ignorable, the classical properties of
a linear medium (that may be intrinsically magneto-electric) can
be parameterised in terms of a constitutive tensor on spacetime
whose properties can in principle be determined by experiments in
non-inertial (accelerating) frames and in the presence of weak but
variable gravitational fields.

Because the properties of the electromagnetic field are so well
understood much of the experimental information  that is collected
about nature  is mediated by this  field. The predictions of pure
quantum electrodynamics have been experimentally verified to high
levels of accuracy and any departures from these predictions are
routinely ascribed to the effects of other interactions. The
classical laws of electrodynamics are also routinely extrapolated
to describe astrophysical phenomena where matter can exist under
extreme conditions of temperature, pressure and density. Even on
terrestrial scales new states of matter and new materials are
regularly being fabricated with surprising electromagnetic
properties that are leading to new technological developments in
communications and nano-science.

To fully understand  phenomena that are induced in bulk matter by
electromagnetism  on these various scales it is necessary to
describe cooperative effects induced by the electromagnetic
interaction. It is, however, difficult to account fully for such
effects in terms of fundamental interactions between charged
particles and photons. The versatility of Maxwell's
phenomenological field equations owes much to the fact that
electromagnetic sources  due to cooperative effects can be
accommodated by using a broad range of constitutive relations
involving a variety of response functions that  account for
polarisation, magnetisation, hysteresis and dispersion in material
continua. Although in principle such response functions can be
calculated in terms of the underlying quantum structure of matter
in many practical situations the constitutive relations are
effectively deduced from experimental data.

In order that this procedure achieves more than a convenient
parametrisation of a particular set of laboratory observations it
is important to embed the response data into a  coherent framework
that respects the basic principles cherished by physical science.
For classical electrodynamics these include the principle that the
underlying theory provides relations between tensor (and spinor)
fields on a four-dimensional spacetime equipped with a light-cone
structure and a pseudo-Riemannian geometry responsible for the
effects ascribed to gravitation.  Such a formulation ensures that
the results of observation  on arbitrarily moving continua by
observers in arbitrary motion (in the presence of an arbitrary
gravitational field) are compatible with the established tenets of
relativistic field theory \cite{hehlbook}. The implementation of
this program is non-trivial particularly when thermodynamic
constraints are included for deformable media.

Some way towards this goal is offered by (covariant) averaging
methods \cite{degrootbook}, \cite{degroot}. These however yield
non-symmetric \SEM tensors for \EM fields in simple media. If the
total \SEM is to remain symmetric this implies that other
asymmetric contributions must compensate and no guidance is
offered to account for such material induced asymmetries. The need
for a symmetric total \SEM tensor is often attributed to
conservation of total angular  momentum despite the fact that such
global conservation laws may not exist in arbitrary gravitational
fields. Although the magnitude of gravitational interactions may
be totally insignificant compared with the scale of those due to
electromagnetism, gravity does have relevance in establishing the
general framework (via the geometry of spacetime) for classical
relativistic  field theory and in particular this framework
offers the most cogent means to define the total \SEM tensor as
the source of relativistic gravitation. This in turn may be
related to a variational formulation
of the fully coupled field system of equations that underpin the
classical description of interacting matter in terms of tensor
(and spinor) fields on spacetime.

\section{Constitutive Relations}

Maxwell's equations for an electromagnetic field in an arbitrary
medium can be written
\begin{align}
d\,F=0 \qquadand d\,\star\, G =j \label{MAXWELL}
\end{align}
where $F$ is the Maxwell 2-form, $G$ is the excitation 2-form and
$j$ is the 3-form electric current source\footnote{The Hodge map
$\star$ is associated with the metric tensor $g$ of spacetime, $d$
denotes the exterior derivative and $i_X$ is the contraction
operator associated with the vector $X$. Details of this notation
can be found \cite{rwt}. All tensors in this article have
dimensions constructed from the SI dimensions $[M], [L], [T], [Q]$
where $[Q]$ has the unit of the Coulomb in the MKS system. We
adopt $[g]=[L^2], [G]=[j]=[Q],\,[F]=[Q]/\ee$ where the
permittivity of free space $\epsilon_0$ has the dimensions $ [
Q^2\,T^2 M^{-1}\,L^{-3}] $ and $c$ denotes the speed of light in
vacuo. }. In general, the effects of gravitation and
electromagnetism on matter are encoded in this system in $\star\,
G$ and $j$.  This dependence may be non-linear and non-local. To
close this system, ``\EM constitutive relations'' relating $G$ and
$j$ to $F$ are necessary. In the following the medium will be
considered as containing polarisable (both electrically and
magnetically) matter with $G$ restricted to a real point-wise
linear function of $F$, thereby ignoring losses and spatial and
temporal material dispersion in all frames {\footnote{ The
electric current $3-$form $j$ will be assumed to describe (mobile)
electric charge and plays no role in subsequent discussions.}}.
Covariant dispersion in linear  material has been discussed in the
optics limit in \cite{dispersion} and the role of the spacetime
metric in the constitutive formulation of Maxwell's theory has
been analysed in \cite{delp}.


The $1-$form electric field $\Me$ and  $1-$form magnetic induction
field $\Mb$  associated with $F$ are defined with respect to an
arbitrary {\it unit} future-pointing timelike $4-$velocity vector
field $U$ by
\begin{align}
\Me = \iuF \qquadand \cc\Mb = \iustarF
\,.
\label{MAXWELL1}
\end{align}
Since $g(U,U)=-1$,
\begin{align*}
F=\Me\wedge \dualu - \star\,(\cc\Mb\wedge \dualu)
\,.
\end{align*}
The field $U$ may be used to describe an {\it observer frame} on
spacetime and its integral curves model idealised  observers.
Likewise the displacement field $\Md$ and the magnetic field $\Mh$
associated with $G$ are defined with respect to $U$ by
\begin{align}
\Md = \iu G \qquadand \Mh/\cc = \iu\star G\,.
\label{MAXWELL2}
\end{align}
Thus
\begin{align*}
G=\Md\wedge \dualu - \star\,((\Mh/\cc)\wedge \dualu)
\,.
\end{align*}
It will be assumed that a material medium has associated with it a
future-pointing timelike unit vector field $V$  which may be
identified with the  bulk $4-$velocity field of the medium in
spacetime. Integral curves of $V$ define the averaged world-lines
of identifiable constituents of the medium. A {\it comoving
observer frame with $4-$velocity $U$} will have $U=V$.

In general $G$ may be a functional of $F$ and properties of the
medium\footnote{ e.g. electrostriction and magnetostriction arise
from the dependence of $ {\cal Z} $ on the elastic deformation
tensor of the medium
.}.
\begin{align}\nonumber
G = {\cal Z}[F,\ldots] 
\,.
\end{align}
Such a functional induces, in general, non-linear and non-local
relations between the fields  $\Md, \Mh$ and $\Me, \Mb$. For
general {\it linear continua}  one may have, for some positive
integer $N$ and collection of {\it constitutive tensor fields}
 $Z^{\,(r)}$ on spacetime, the relation
\begin{align}\nonumber
G = \Sigma_{r=0}^{N}{ Z^{\,{(r)}}}(\nabla^{\,r} F,\ldots)
\end{align}
in terms of some spacetime connection $\nabla$. Additional
arguments refer to  variables independent of $F$ and its
derivatives. For the non-dispersive  linear media under
consideration here, we restrict to
\begin{align}
G=Z(F) \label{Media_General_Constitutive}
\end{align}
 for some {constitutive tensor field} $Z$. In the
vacuum $G=\epsilon_0 F$.

 A particularly simple linear isotropic but inhomogeneous medium may be described by a
bulk $4-$velocity field  $V$, a relative permittivity  scalar
field $\epsilon$ and a non-vanishing  relative permeability scalar
field  $\mu$. To a comoving observer ($U=V$)  with   a history
that coincides with one of the integral curves of $V$ the local
constitutive relations become
\begin{align}\nonumber
\Md = \ee\epsilon\, \Me \qquadand \Mh = (\mu_0\mu)^{-1} \Mb \,
.\end{align}

For a non-magneto-electric but anisotropic medium, the relative
permittivity $\epsilon$ and inverse  relative permeability
$\mu^{-1}$ become {\it spatial tensor fields} on spacetime. More
generally, the electromagnetic fields  measured by a co-moving
observer may be related by
\begin{align}
\Md = \PermDE(\Me) + \PermDB(\Mb) \qquadand \Mh = \PermHE(\Me) +
\PermHB(\Mb)
\label{Media_Constitutive_dh}
\end{align}
where $\PermDE,\PermDB,\PermHE,\PermHB$ are spatial tensors. From
(\ref{MAXWELL1}), (\ref{MAXWELL2}),
(\ref{Media_General_Constitutive}) it follows that this
constitutive relation may be expressed covariantly as
\begin{align}\nonumber
\PermTen(F) = \PermDE(\ivF)\!\wedge\!\dualv +
\PermDB(\ivstarF)\!\wedge\!\dualv - \star(\PermHE(\ivF)\!\wedge\!\dualv) -
\star(\PermHB(\ivstarF)\!\wedge\!\dualv)
\,.
\label{Media_Z_decomp}
\end{align}
If $\PermHE=\PermDB=0$ then $\PermDE=\ee\epsilon$ and
$\PermHB=(\mu_0\mu)^{-1}$.  However, for such materials one cannot
assert that $\PermHE, \PermDB$ remain zero in all frames. Materials
with the general constitutive relation (\ref{Media_Constitutive_dh})
are often referred to as {\it magneto-electric} \cite{post},
\cite{odell}.


Maxwell's equations (\ref{MAXWELL}) in a medium in spacetime $M$,
with $j=0$, follow naturally as a local extremum of the action
functional
  $S[A,g]=\int_M \Lambda$ under $A$ variations, where
locally
   $F=d\,A$, $G=\PermTen(F)$,  and
\begin{equation}\nonumber
\cc\Lambda = \tfrac{1}{2}\,F \wedge \star\, G = \tfrac12\,F
\wedge \star\, \PermTen(F) \, 
\end{equation}
provided the $4-$th rank tensor $Z$ is taken to be self-adjoint:
$Z^{abcd}=Z^{cdab}$.

The dependence of this action on the metric resides in the $\star$
map and $Z$. Thus the form of the variational derivative of this
action under metric perturbations will depend on the response of
the constitutive tensor $Z$ to gravitation. The above form of the
constitutive relations relating electric and magnetic fields in an
arbitrary timelike frame  yields a natural tensor relation between
$Z$ and $\tilde V\equiv g(V,-)$ and so offers a natural dependence
of $Z$ on $g$ for inhomogeneous, anisotropic (non-dispersive)
magneto-electric continua.

For such a $Z$  one may compute by a metric variation of  the
above action  the stress-energy-momentum tensor associated with
the electromagnetic field in such a medium. After a non-trivial
calculation one obtains:
\begin{align}
T= \tfrac12 \Big(i_a F\otimes i^a G +  i_a G\otimes i^a F -
\star(F\wedge\star G) g +  \dualv \otimes \GenPoyn + \GenPoyn
\otimes \dualv \Big)
\label{intro_Abraham_Stress_T}
\end{align}
where the 1-form
$\GenPoyn=\star\big(\ivF\wedge\ivstarG\wedge\dualv +
\ivstarF\wedge\ivG\wedge\dualv\big).$

This  reduces in gravity-free Minkowski spacetime to the tensor
attributed historically to Abraham. The details of the derivation
in the considerably wider context outlined above may be consulted
in \cite{calc}. In terms of comoving fields, defined by
(\ref{MAXWELL1}), (\ref{MAXWELL2}) with $U=V$,
(\ref{intro_Abraham_Stress_T})
 may
be written in the manifestly symmetric form:
\begin{equation}\nonumber
\begin{aligned}
T =& -\tfrac{1}{2} (\Me \otimes \Md + \Md \otimes \Me)
-\tfrac{1}{2} (\Mh \otimes \Mb + \Mb \otimes \Mh)
\\ &+
\tfrac{1}{2} ( g(\tilde{\Me},\tilde{\Md}) +
g(\tilde{\Mh},\tilde{\Mb})) ( g + 2 \dualv \otimes \dualv ) +
(\dualv \otimes \tilde{S} + \tilde{S} \otimes{V})
\end{aligned}
\end{equation}
where the Poynting 1-form
$\tilde{S} = \star( \dualv \wedge \Me \wedge \Mh).$

If, by contrast  $\PermTen$ is chosen to be  totally {\it
independent of the metric}  and hence $\tilde V$,   the resulting
stress-energy-momentum tensor becomes
\begin{align}\nonumber
T= \tfrac12 i_a G\otimes i^a F  + \tfrac12 i_a F\otimes i^a G -
\tfrac12 \star(F\wedge\star\, G) g
\end{align}
showing clearly its independence of the 4-velocity $V$  of the
medium. In the absence of gravity such a tensor reduces to that
obtained by symmetrising the one proposed by Minkowski.

\section{Conclusion}

There has been a rapid development in recent years in the
construction of \lq\lq traps" for confining collective states of
matter on scales intermediate between macro- and micro-dimensions.
Condensates of cold atoms and fabricated nano-structures offer
many new avenues for technological development when coupled to
probes by \EM fields. The constitutive properties of such novel
material will play an important role in this development. Space
science is also progressing rapidly and can provide  new
laboratory environments with variable gravitation and controlled
acceleration in which the properties of such states of matter may
be explored. {\it Our results offer a new and efficient way to
establish a coherent understanding of the stresses and
energy-momentum exchanges induced by electromagnetic interactions
with such matter in terms of a fully relativistic covariant
variational framework.} Supplemented with additional data based on
mechanical and elasto-dynamic responses one thereby gains a more
confident picture of a total phenomenological symmetric \SEM
tensor for  a wide class of moving media than that based on
previous ad-hoc choices.

\vspace{0.5cm} \noindent {\bf Acknowledgements} The authors are
grateful D. Burton and A. Noble for helpful discussions, the EPSRC
and  Framework 6 (FP6-2003-NEST-A) for financial support for this
research and the support provided by the Cockcroft Institute, UK.



\end{document}